\documentclass{PoS}

\title{Seed methods for linear equations in lattice qcd problems with multiple right-hand sides }

\ShortTitle{Seed methods for linear equations in lattice qcd}

\author{\speaker{Abdou Abdel-Rehim}
        \thanks {New address: Department of Physics and Department of Computer Science, The College of William and Mary.}\\
        Department of Physics, Baylor University, Waco, TX 76798-7316, U.S.A.\\
        E-mail: \email{amrehim@cs.wm.edu}}

\author{Ronald B. Morgan\\
        Department of Mathematics, Baylor University, Waco, TX 76798-7328, U.S.A.\\
        E-mail: \email{Ronald$\_$Morgan@baylor.edu}}

\author{Walter Wilcox\\
        Department of Physics, Baylor University, Waco, TX 76798-7316, U.S.A.\\
        E-mail: \email{Walter$\_$Wilcox@baylor.edu}}

\abstract{We consider three improvements to seed methods for Hermitian linear systems with multiple right-hand 
          sides: only the Krylov subspace for the first system is used for seeding subsequent right-hand 
          sides, the first right-hand side is solved past convergence, and periodic re-orthogonalization 
          is used in order to control roundoff errors associated with the Conjugate Gradient algorithm. The method 
          is tested for the case of Wilson fermions near kappa critical and a considerable speed up in the convergence 
          is observed.}

\FullConference{The XXVI International Symposium on Lattice Field Theory \\
         July 14 - 19, 2008\\
         Williamsburg, Virginia, USA}

\begin{document}

\section{Introduction}
Solving sparse linear systems of equations with many right-hand sides has important applications in lattice QCD. One
particular approach that has been found to play a key role as one approaches light quark masses is deflation of small
eigenvalues(see \cite{Deflation}-\cite{LanDR} for a description of different deflation techniques). Deflation techniques rely on computing
eigenvectors with small eigenvalues and projecting over them to speed up the convergence for extra right-hand sides.
The eigenvectors are computed either separately or concurrently while solving a subset of the right-hand sides. Here 
we consider seed methods \cite{smith}-\cite{chan} for Hermitian positive definite systems. The basic idea of Seed methods is to choose
one right-hand side as a ``seed system'' and use the Krylov subspace developed for the seed system as a projection space
for the remaining right-hand sides. At the end of solving the seed system, one has developed a better initial guess for
the remaining systems. At that point, one has the choice of either repeating the seeding by choosing the second right-hand 
now as a seed system or solving the remaining systems using the Conjugate Gradient algorithm and the new initial guesses.
Traditionally, the seeding is repeated and the efficiency of the method depends on how closely related the right-hand
sides are. 

Since deflating small eigenvalues seems to be the dominant factor in speeding up the solution of
subsequent right-hand sides, we propose three improvements to the standard seed method that helps achieve that. The main idea
is to make sure that the Krylov subspace used in seeding contains good approximations of those important eigenvectors, even though
we are not solving explicitly for them. First, we seed only once using the first right-hand side. Second, we run CG for the first 
right-hand side past the convergence limit. Third, since roundoff errors will affect the CG algorithm we have to apply some form 
of a reorthogonalization step. In this work we use periodic reorthogonalization. Because of the need for the reorthogonalization 
step, one has to save all the Krylov vectors during the solution of the first right-hand side. The extra memory and 
reorthogonalization cost has to be taken into consideration as will be discussed (see \cite{iseedcg} for a more detailed discussion of 
the method). In section \ref{algorithm} we describe the Improved Seed-CG
algorithm. Results from testing the algorithm with a lattice QCD example are given in section \ref{results}.

\section{Improved Seed-CG Algorithm}
\label{algorithm}
Consider the linear systems $Mx^j=b^j \quad ;j=1,2,...,nrhs$, where $M$ is a symmetric (Hermitian) positive definite matrix. Let $x_0^j$ 
be the initial guesses and $r_0^j=b^j-Mx_0^j$ the initial residual vectors. In a seed
approach, we'll solve the first right-hand side using the Lanczos (CG) algorithm to build a Krylov subspace 
${\cal{K}}=span\{r_0^1, Mr_0^1, M^2r_0^1,..,M^{m-1}r_0^1\}$ whose orthonormal bases is
$V=\{v_1,v_2,..,v_m\}$ where $m$ is the dimension of the Krylov subspace at which the first right-hand side has converged to the desired
accuracy. The subspace developed for the first right-hand side is used to obtain a better initial guess for the remaining right-hand
sides using a Glerkin projection such that $x^j ;j=2,3,..,nrhs$ are the improved initial guesses. A faster convergence of 
the extra right-hand sides is obtained by using these improved initial guesses with a standard CG solver. If $T$ is the tridiagonal matrix
obtained in the Lanczos algorithm while solving the first right-hand side, then after the first right-hand side is solved we have:
\begin{equation}
 x^j  =  x_0^j + Vy^j ; \quad y^j=T^{-1}(V^\dagger r_0^j) ; \quad \quad j=1,2,..,nrhs.  
\end{equation} 

In order to update $x^j$ at every iteration we use LU factorization of the Lanczos tridiagonal matrix (in case of indefinite
systems a QR or LQ factorization should be used). The LU factorization is given by (for $m=5$ as an example):

\begin{equation}
T=\left(
  \begin{array}{c c c c c}
   \alpha_1 & \beta_1 &           &         &       \\
   \beta_1  & \alpha_2 & \beta_2  &         &       \\
            & \beta_2  & \alpha_3 & \beta_3 &       \\
            &          & \beta_3  & \alpha_4&\beta_4\\       
            &          &          & \beta_4 &\alpha_5\\
  \end{array}
\right)
=
\left(
  \begin{array}{c c c c c}
   1       &  &  &  & \\
   \gamma_1& 1 &  &  & \\
           & \gamma_2 & 1 &  & \\
           &          & \gamma_3 & 1 & \\       
           &          &          & \gamma_4 & 1\\
  \end{array}
\right)
\times
\left(
  \begin{array}{c c c c c}
   \delta_1 & \beta_1 &  &  & \\
   & \delta_2 &\beta_2  &  & \\
           &  & \delta_3 & \beta_3 & \\
           &          &  & \delta_4 & \beta_4\\       
           &          &          &  & \delta_5\\
  \end{array}
\right).
\end{equation}
Using this factorization the small linear systems $Ty^j=V^\dagger r_0^j$ are solved in an incremental way 
during every iteration. The procedure is similar to what is used with the D-Lanczos algorithm \cite{saadbook}. 

The vectors built with Lanczos three-term recurrence lose orthogonality as the dimension of the subspace
becomes large enough for some of the eigenvectors to begin to converge \cite{paige},\cite{matalg}. One way to 
fix this problem is to perform an extra reorthogonalization step. Here, we implement a periodic 
reorthogonalization ({\it{PO}}) \cite{po},\cite{matalg}. 
One chooses a frequency $k$ of performing the {\it{PO}} step such that during iteration number $i$, if $i$
is a multiple of $k$, we reorthonormalize $v_{i}$ against all vectors  $v_1,v_2,...,v_{i-1}$ and reorthogonalize
$\hat{v}_{i+1}$ against all vectors $v_1,v_2,...,v_{i}$ where $\hat{v}_{i+1}$ is the un-normalized $i+1$ vector 
\cite{thick_restart},\cite{matalg}. The Lanczos version of single
seed CG with periodic reorthogonalization is described below.

\vspace{0.2 in}
\begin{tabular}{l}
\hline
{\bf{Algorithm 1}}: Seed-CG(m,k) \\
\hline
{\bf{1. Start:}} Choose $m$, the maximum size of the subspace, and $k$, the frequency\\ 
\quad of reorthogonalization. The number of right-hand sides is $nrhs$. For $j=1,2,..,nrhs$,\\
\quad set the approximate solution denoted by $x^j$ equal to the initial guess (which maybe \\
\quad the zero vector) and compute the initial residual $r_0^j$.\\

{\bf{2. First Lanczos iteration:}} $\beta_0=||r_0^1||;\quad v_1=r_0^1/\beta_0;$ \\
$ \quad  f=Mv_1;\quad \alpha_1=v_1^\dagger f;\quad f=f-\alpha_1v_1;\quad \beta_1=||f||$.\\

{\bf{3.Linear equations for first iteration:}} $\delta_1=\alpha_1;\quad w_1=v_1/\delta_1$;\\
$\quad \zeta_1=\beta_0 ; \quad x^1 = x^1 + \zeta_1 w_1$. \\

{\bf{4. Other right-hand sides for first iteration:}} For $j=2,..,nrhs$, $\eta^j=v_1^\dagger r_0^j $;\\
$\quad  x^j = x^j + \eta^j w_1$. Set $i=2$.\\

{\bf{5. Lanczos iteration:}}$f=Mv_i-\beta_{i-1}v_{i-1}$; $\alpha_i=v_i^\dagger f$; $f=f-\alpha_i v_i$.\\  

{\bf{6. Linear equations:}}$\gamma_{i-1}=\beta_{i-1}/\delta_{i-1}; \quad \delta_i=\alpha_i-\gamma_{i-1}\beta_{i-1};$\\
$\quad  w_i=(v_i-\beta_{i-1}w_{i-1})/\delta_i;\quad \zeta_i=-\gamma_{i-1}\zeta_{i-1};\quad x^1=x^1+\zeta_iw_i.$\\

{\bf{7. Other right-hand sides:}} For $j=2,3,..,nrhs$ , $\eta^j=v_i^\dagger r_0^j -\gamma_{i-1}\eta^j; \quad 
x^j = x^j + \eta^j w_i$.\\

{\bf{8. Reorthogonalization:}} If $i$ is a multiple of $k$, then reorthonormalize $v_i$ against\\ 
\quad $v_1,v_2,...,v_{i-1}$ and reorhtogonalize $f$ against $v_1,v_2,..,v_i$.\\

{\bf{9. Finish iteration and go to next iteration:}} $\beta_i=||f||; v_{i+1}=f/\beta_i$. \\
\quad if $i < m$, set $i=i+1$ and go to step $5$.\\
\hline\\
\end{tabular}
\vspace{0.2 in}
 
The value of $m$ is chosen in such a way that the first system is solved past the convergence limit that we 
require for the subsequent systems. This is to ensure that the Krylov subspace is large enough to contain 
accurate eigenvectors with small eigenvalues. The reorthogonalization frequency is chosen to balance 
the cost of reorthognalization during the solution of the first right-hand side and the gain we get for the 
subsequent systems. In most cases we have many right-hand sides to solve so this reorthogonalization
cost is not significant.

\section{A Lattice QCD Example}
\label{results}
\begin{figure}
\begin{center}
\includegraphics[width=0.7\textwidth]{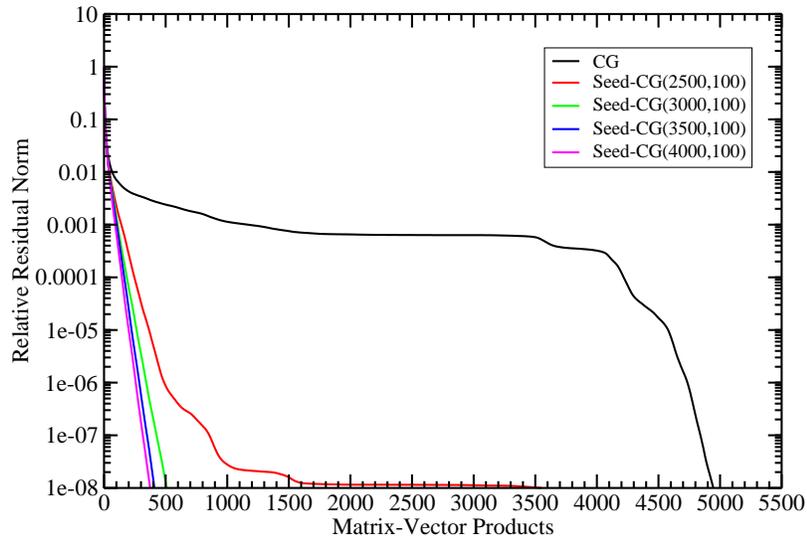}
\vspace{0.1 in}
\caption{Results for the second right-hand side with seed-CG(m,100) on a $20^3\times 32$
quenched Wilson configuration at $\beta=6.0$ and $\kappa=0.15720$. Standard
CG results are shown for comparison.}
\label{figure1}
\end{center}
\end{figure}

\begin{figure}
\begin{center}
\includegraphics[width =0.7\textwidth]{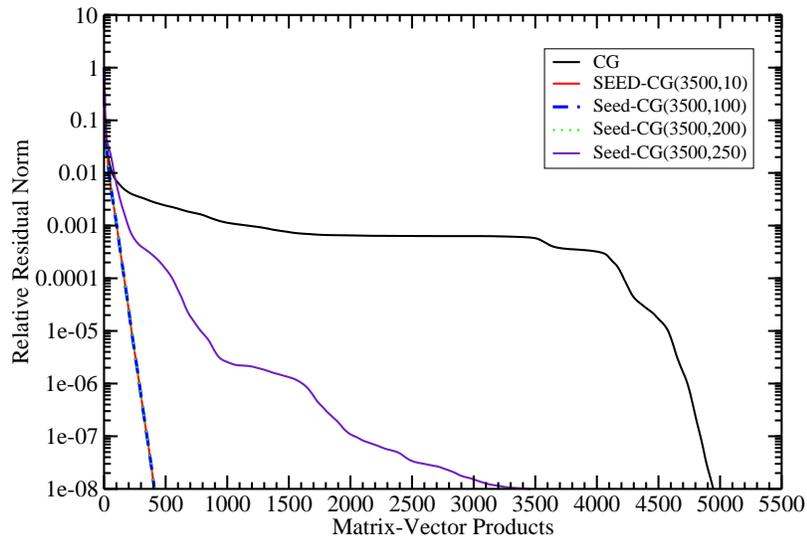}
\vspace{0.1 in}
\caption {Effect of changing the frequency of reorthogonalization with  m=3500 for the second right-hand side 
on the same configuration. Standard CG results are shown for comparison.}
\label{figure2}
\end{center}
\end{figure}

The method is tested for Wilson fermions on a $20^3\times32$ quenched Wilson
configuration at $\beta=6.0$ . We choose $\kappa$ to be essentially 
$\kappa_{critical}\approx 0.15720$ to make the problem difficult. Since we seed only once, 
we need to consider only two sample right-hand sides. These were taken to be point sources 
with Dirac and color indices $1,1$ and $1,2$ respectively. We solve the even-odd
preconditioned systems $Ax^j=\phi^j; \quad j=1,2$ with $A$ non-Hermitian.
We convert this problem into a Hermitian positive definite problem as $A^\dagger A x^j = A^\dagger \phi^j$.
However, the residual norm that will be monitored for convergence will always be the one that 
corresponds to the original non-Hermitian system.The relative residual norm at convergence is chosen to be
$10^{-8}$. Standard CG for this problem requires Krylov subspace of
dimension about 2,500 vectors (5000 matrix-vector products). In order to
reduce the matrix-vector products for the second right-hand side it will be
necessary to solve the first right-hand side past convergence. In Figure \ref{figure1},
results for the second right-hand side is shown when the first right-hand side
is solved using Krylov subspaces with dimensions 2500, 3000, 3500, and 4000. For this test,
we used $k=100$ for the reorthogonalization frequency. Finding the largest
value of $k$ that could be used for this problem was done by experimenting
with different choices of $k$ values. In Figure \ref{figure2}, we show results for the second
right-hand side using a subspace dimension 3500 and $k$ values 10, 100, 200,
and 250. As can be seen from Figure \ref{figure2}, a $k$ value less than 250 is suitable for
this problem. Having a large $k$ value reduces the cost associated with reorthogonalization.

The results for the second and additional right-hand sides shows a considerable gain in terms of the number of matrix-vector
products of CG when using the initial guesses from the seeding. For the first right-hand side, there is additional overhead
as compared to standard CG that comes from three sources: the need to store all Krylov subspace vectors, reorthogonalization,
and the need to solve past the convergence level. There is also the cost of projecting over for subsequent 
right-hand sides (seeding), however, in
our example we have only one additional right-hand side and this cost won't be discussed. The memory cost 
can be tolerated given that it is only needed during the solution of the first right-hand side and should be allocated and 
deallocated at run time. This pays off, particularly when we have many right-hand sides. In the above example, the results were 
obtained using the high-performance computer cluster at Baylor University which has 8 processors 
per node at 2.66GHZ and 16GB RAM per node. The run was done on 50 processors using 5 processors per node allowing for 3.2 GB RAM per process. 
It was possible to allocate up to 4000 Krylov vectors (every vector is distributed over the 50 processes) using this configuration without 
a need for memory swapping that leads to a slower run. The second overhead comes from the reorthogonalization step. The value
of the reorthognalization frequency has to be found by experimenting in order to find the optimal value that reduces the reorthogonalization
cost and at the same time increases the gain for matrix-vector products of subsequent right-hand sides. Using the same computing configuration
as described above we made a simple timing test for the cost of reorthogonalization. The results are shown in Table \ref{timing}.
\begin{table}
\begin{center}
\begin{tabular}{|c|c|}
\hline
Frequency of reorthogonalization & time in CPU seconds \\
\hline
10 & 1253 \\
100& 598 \\
200& 320 \\
250& 305 \\
no reorthogonalization & 227 \\
\hline
\end{tabular}
\caption{Timings for solving the first right-hand side with different reorthogonalization choices.}
\label{timing}
\end{center}
\end{table}
Since values of reorthogonalization frequencies 10, 100, and 200 seem to give the same improvement for the second
right-hand side as can be seen in Figure \ref{figure2}, we can choose $k=200$ and in this case the extra cost of
the reorthogonalization step is less than 2 times the cost of no reorthognalization. The final overhead to consider is the
need to solve the first right-hand side past convergence. We found that in terms of matrix-vector products, this leads at most to
a factor of 2 also. In our example, as can be seen from Figure \ref{figure1}, solving the first right-hand side using 3000, 3500, and 4000
vectors gives similar gains for the second right-hand side using 1000, 2000, and 3000 extra matrix-vector products respectively (standard CG
for the first right-hand side uses $\approx 5000$ matrix-vector products). So, this leads to a factor of  $20\%$, $40\%$, and $60\%$ 
extra matrix-vector products.
So, the extra cost of solving the first right-hand side associated with the reorthogonalization and the need to go past convergence
leads to about factor of 2 times the cost of standard CG. This extra cost is acceptable given the gain we get when we have many right-hand sides.  

\section*{acknowledgements}
Calculations were done at the HPC systems at Baylor University. A. Abdel-Rehim would like to thank
Departments of Physics and Computer Science at the College of William and Mary as well as Thomas Jefferson
National accelerator facility for support during the writing of this proceedings report.

\end{document}